# A curated global dataset of open science initiatives and policies

Abdelghani Maddi[1]*, Marc-André Simard[2] and Isabelle Dorsch[1]

[1] Sorbonne Université, CNRS, Groupe d'Étude des Méthodes de l'Analyse Sociologique de la Sorbonne, GEMASS, F-75017 Paris, France.

[2] École de bibliothéconomie et des sciences de l'information, Université de Montréal, Canada.

*Correspondence: abdelghani.maddi@cnrs.fr; 59/61 rue Pouchet, 75017 Paris, France.

ORCID IDs: Isabelle Dorsch 0000-0001-7391-5189; Marc-André Simard 0000-0003-3795-0053; Abdelghani Maddi 0000-0001-9268-8022.

## I. ARTICLE AND DATASET INFORMATION

### Abstract

Inventories of open science initiatives differ in scope, geographic coverage and classification, making it difficult to compare infrastructures, services, networks and policies. We developed COSMI Hub within the OPENIT project to bring these sources together in a curated dataset. The release archived on 8 October 2026 contains 1,186 records and 71 variables covering initiative identity, source provenance, geography, record level, operational status, thematic activities, governance, funding and external identifiers. We identified candidates through structured directories, policy documents, official websites and suggestions from the open science community, then checked identities, reconciled duplicates and assessed supporting evidence. The dataset distinguishes standalone initiatives, services, deployments, umbrella organisations and policy instruments. Records can be assigned several thematic dimensions, with uncertainty retained where evidence is incomplete. We describe the collection and coding methods, the data structure and file-level checks, together with considerations for reuse. The Zenodo release is fixed, whereas the companion dashboard is updated separately. Because we assembled the collection purposively, it cannot be used to estimate the worldwide number of open science initiatives.

### Résumé (français)

Les répertoires d'initiatives de science ouverte diffèrent par leur périmètre, leur couverture géographique et leurs catégories, ce qui complique la comparaison des infrastructures, services, réseaux et politiques. Dans le cadre du projet OPENIT, nous avons développé COSMI Hub pour rapprocher ces sources dans un jeu de données harmonisé. La version archivée sur Zenodo le 8 octobre 2026 contient 1 186 notices et 71 variables portant sur l'identité des initiatives, la provenance des informations, leur localisation, leur statut, leurs activités, leur gouvernance, leur financement et certains identifiants externes. Nous avons identifié des notices candidates à partir de répertoires, de documents de politique publique, de sites officiels et de suggestions de la communauté, puis vérifié leur identité, rapproché les doublons et examiné les sources justificatives. Le corpus distingue les initiatives autonomes des services, déploiements, organisations faîtières et instruments de politique publique. Une notice peut relever de plusieurs dimensions de la science ouverte ; l'incertitude est conservée lorsque les sources ne permettent pas de trancher. L'article présente la méthode de constitution et de codage, la structure du fichier, les contrôles réalisés et les précautions de réutilisation. La version Zenodo est figée, tandis que le tableau de bord est actualisé séparément. Constitué par collecte raisonnée, ce corpus ne permet pas d'estimer le nombre total d'initiatives de science ouverte dans le monde.

**Dataset metadata**

Dataset title: COSMI Hub: A Curated and Evolving Dataset of Open Science Initiatives Worldwide.

Dataset creators: Isabelle Dorsch; Marc-André Simard; Abdelghani Maddi.

Archive / publisher: Zenodo; public access; resource type: dataset.

Published release: 8 October 2026; dataset version 1.0.0.

Persistent identifier: https://doi.org/10.5281/zenodo.23233124

File: cosmihub_master_2026-10-08.csv; one UTF-8 CSV (approximately 1.4 MB); 1,186 records and 71 variables.

Licence for the data: Creative Commons Attribution 4.0 International (CC BY 4.0).

File integrity: MD5 ca754336562d01eb63e16814afd086fe (Zenodo deposit).

Language and coverage: coding and field names in English; the documented initiatives and source pages cover several countries and languages.

Dataset contact: Abdelghani Maddi, corresponding author (abdelghani.maddi@cnrs.fr).

Related publication: Dorsch, Simard, and Maddi (2025), COSMI Hub conference contribution, https://doi.org/10.5281/zenodo.17305636.

Interactive companion: https://amcm.shinyapps.io/openit/ ; software: https://github.com/abdelghani-maddi/osinit.

## II. ARTICLE TEXT

### 1. Rationale and scope

Inventories of open science initiatives serve different purposes. Infrastructure directories list platforms and services, monitoring inventories describe efforts to measure open science activity, and policy reports document instruments within particular jurisdictions. Their entries use different units and categories. To compare them, we needed a common structure and traceable sources.

We developed COSMI Hub (Community Open Science Mapping Initiatives Hub) as part of the ANR-funded OPENIT project (https://www.gemass.fr/contract/openit/). The collection grew out of ASAPbio's prototype directory (Corker & Shaw, 2024) and expanded beyond open access initiatives to cover a wider range of open science activities (Dorsch et al., 2025). Here we describe the CSV snapshot archived on 8 October 2026 (Dorsch et al., 2026; https://doi.org/10.5281/zenodo.23233124). Our R Shiny dashboard

(https://amcm.shinyapps.io/openit/) draws on an evolving working database and may therefore show different totals. We reviewed and annotated the COSMI Hub records rather than copying entries unchanged from the source directories.

COSMI Hub can be used to study open science infrastructures, governance, policy and geography, including questions in library and information science and scientometrics. Its fields support descriptive comparisons and independent recoding. Because we did not draw a probability sample, however, it cannot be used to estimate how many open science initiatives exist worldwide.

## 2. Data collection and curation

### 2.1. Observation unit and inclusion criteria

Each row in COSMI Hub describes a documented object. The variable record_level specifies its granularity (standalone initiative, service, deployment or umbrella organisation), whereas object_type_openit captures its functional type. These are separate attributes: a service or umbrella organisation is not necessarily dependent in legal or organisational terms. The variable initiative_family links related records without merging them into a single observation. For example, OpenAIRE Monitor: Aurora Universities and OpenAIRE Monitor: EUt+ are distinct deployment records sharing the initiative_family value OpenAIRE Monitor. We retain a platform and one of its named services as separate records when each has a documented remit; likewise, we record distinct policy instruments separately rather than merging them into a single country entry. During screening, we checked whether a candidate represented a nameable initiative, service, deployment, umbrella organisation or identifiable policy instrument; whether an official website, attributable directory entry or authoritative policy document supported its identity and open-science remit; and whether its name, purpose and available identifiers distinguished it from existing records. Candidates with insufficient or conflicting evidence remained in working audit lists rather than being treated as confirmed records. Planned or discontinued initiatives could be included when their identity and status were documented. The dataset is a cross-sectional, non-probability collection of records, not a list of independent legal entities.

A parent network can operate several named services, and an organisation already listed as an initiative may also run a monitoring instrument. These links matter for counting records: treating every row as an independent organisation would count some related entities more than once. We used country profiles and umbrella homepages as supporting evidence when they did not document a distinct initiative or policy instrument.

### 2.2. Source selection and candidate identification

We started with 96 entries from ASAPbio's prototype directory (Corker & Shaw, 2024), then sought candidates in other structured inventories, official websites and documentary sources, as well as through suggestions from community members and requests for corrections. These inputs were screened against the same inclusion criteria as directory records, not imported automatically. The archived CSV does not include a separate log of community suggestions or correction requests. An entry in another directory was not sufficient for inclusion on its own.

We used Infra Finder, developed by Invest in Open Infrastructure (IOI), to identify and check infrastructure and service records. Its underlying data are deposited on Zenodo (Invest in Open Infrastructure, 2026), and Skinner et al. (2024) describe the tool. We also drew candidates from the monitoring-focused inventory of OSMI Working Group 2 (OSMI WG2, 2026), which has a narrower scope than COSMI Hub. We also consulted Open Research Beacon (ICOR, 2025), UNESCO Open Science Initiatives (UNESCO, 2026), GO FAIR Implementation Networks (GO FAIR, n.d.), LIBSENSE (Abbott et al., 2024), INOSC (Eerland et al., 2025), the LA Referencia–UNESCO survey (LA Referencia & UNESCO, 2024) and TSOSI (Larrieu, 2025). Depending on their scope, we used these resources to identify candidates, verify attributes or examine geographic coverage. We did not import every entry from each source.

We used the European Commission report Open access policies in EU Member States and beyond (Spichtinger, 2026) to compile policy records. Our working integration log records 171 distinguishable policy instruments across the 27 EU Member States and eight comparator countries: Brazil, Canada, China, Colombia, Japan, Switzerland, the United Kingdom and the United States. We created separate records for strategies, laws or mandates with distinct names or functions, but not for general country descriptions or implementation bodies. The report describes policies as documented in 2026 and does not capture subsequent changes in their status.

We record provenance in source, source_url and source_record_id where available. A record may cite several sources, so the values in source are not mutually exclusive. Our working source registry distinguishes candidate directories from audit lists and geographic reference materials.

### 2.3. Harmonisation, screening and deduplication

For each candidate, we reviewed names and aliases, official websites, institutional or geographic base, remit, record level, operational status and provenance. We then normalised names and URLs and mapped country values to ISO 3166-1 alpha-3 codes where applicable. International and regional entries use project-defined codes. When evidence was missing or conflicting, we left the relevant fields unresolved.

To identify duplicates, we compared names, official websites, source identifiers and initiative families. When two candidates referred to the same object, we retained a stable internal identifier and combined their source attributions; distinct services and deployments remained separate records. For older source entries without original row identifiers, we assigned locally generated keys marked *_LEGACY, which are not identifiers supplied by the upstream source. Unresolved matches and candidate records remain in our working audit sheets, which are not included in the archived CSV. The release therefore does not contain a complete log of individual review decisions.

### 2.4. Topical categorisation and organisational attributes

To categorise the topical scope of each record, we use the ten dimensions of the revised open science taxonomy (da Silveira et al., 2023), informed by the UNESCO Recommendation on Open Science (UNESCO, 2021). We assess each record against the dimensions using official websites and other attributable documents. A record can receive more than one dimension, and we use 'possible' when the evidence is inconclusive. Table 1 gives the corresponding CSV variable names and labels.

**Table 1. Open science dimensions and corresponding CSV variables.**

| CSV column name | Open-science dimension |
| --- | --- |
| dim_open_access | Open access |
| dim_open_data | Open data |
| dim_open_reproducible_research | Open reproducible research |
| dim_open_responsible_evaluation | Open and responsible evaluation of science |
| dim_policy_guidelines | Policy, declarations and guidelines of open science |
| dim_open_education | Open education |
| dim_open_innovation | Open innovation |
| dim_open_infrastructure_tools | Open Science infrastructure and tools |
| dim_citizen_science_participative | Citizen science, open and participative |
| dim_open_dialogue_knowledge_systems | Open dialogue with other knowledge systems |

We initially worked with a codebook that assigned one primary open science category to each initiative. In subsequent team discussions, we recognised that this could obscure initiatives active in several areas and make the choice of a dominant category subjective. AmeliCA (https://amelica.org/), for example, spans several areas of open science and cannot be adequately represented by a single label. We therefore decided to code all the dimensions supported by documentary evidence rather than force each initiative into one category (Table 1).

Separately from topical categorisation, we record descriptive organisational attributes from documentary sources: type of lead actor, governance arrangements, community involvement, scope, non-profit status, open-source software, multilingual provision and presence in the Global South. These are source-informed attributes, not statistical clusters or the result of a formal thematic-analysis procedure. We distinguish community participation from community ownership rather than treating one as evidence of the other. Separate economic fields describe funding arrangements and, where available, supporting sources and confidence levels.

All three authors contributed to the initial topical classification. The procedure subsequently evolved from assigning a single primary category to identifying multiple evidence-supported dimensions for each record. Some ambiguous cases were discussed informally by the authors, including when comparing alternative classifications. This was not a systematic independent double-coding or formal adjudication exercise; the archived release does not retain the independent coding records needed to calculate an intercoder agreement coefficient.

### 2.5. Identifier enrichment and entity matching

We added optional ROR, Wikidata and TSOSI identifiers to support entity matching. The TSOSI team assisted with an enrichment batch documented in our project records: initiative names and countries were used to search ROR and Wikidata, proposed matches were checked manually, and TSOSI identifiers were linked where supported. The batch comprised 121 ROR identifiers, 388 Wikidata identifiers and 61 TSOSI identifiers. These are figures from the enrichment exercise, not counts of populated cells in the archived CSV. We record the source of identifier enrichment in identifier_enriched_by. A missing external identifier means that no match was recorded, not that the organisation is absent from the external registry.

External identifiers must be interpreted at the correct entity level. A service may have no direct ROR match even when its host institution does, and a Wikidata or TSOSI identifier may refer to an umbrella organisation rather than a particular service. Record-level joins therefore require identity and granularity checks.

### 2.6. Data quality and validation

#### 2.6.1. File integrity and structural checks

We verified the archived CSV's row and column counts, the uniqueness of record identifiers, and the completeness of selected fields. It contains 1,186 rows and 71 columns, with 1,186 unique id values. We found no repeated names after trimming whitespace and comparing case-insensitively. These checks establish structural properties of the deposited CSV; they do not establish the substantive accuracy of each topical, governance or operational-status code, which would require record-by-record comparison with the documentary evidence.

#### 2.6.2. Coverage and missing data

The COSMI Hub v1.0.0 snapshot contains 987 standalone initiatives, 135 services, 32 umbrella entries and 32 deployments. For operational_status, 724 records are unknown, 416 ongoing, 38 discontinued, six pilot/pre-production and two planned. Because most statuses are unknown, comparisons involving ongoing initiatives are sensitive to uneven availability of status information.

Initiative websites are blank for 144 records and for 182 about pages (variable: about_link). Ten records lack latitude and longitude; iso3 is populated for all records. ROR identifiers are missing or marked 'NA' for 1,066 records, Wikidata identifiers for 801 and TSOSI identifiers for 1,125. These counts describe external identifier coverage, not necessarily errors in the records.

The ten topical dimension columns contain mainly 'yes' and 'no' values, with rare 'possible' assignments indicating insufficient evidence for a confident 'yes'; several dimensions may apply to a single record. A 'no' value reflects the available documentation, not proof that an activity is absent. Because systematic independent double-coding and adjudication records are unavailable, intercoder agreement cannot be estimated from this release.

#### 2.6.3. Source review and reproducibility limits

We used source registries, candidate lists and audit worksheets while checking sources, coding records and reconciling identities. These working files were not deposited with the COSMI Hub v1.0.0 CSV archived on Zenodo on 8 October 2026. Readers can reproduce descriptive counts and structural checks directly from the CSV (including row totals, unique identifiers and missing-value counts), but cannot reconstruct every inclusion or coding decision without the separate audit records. Reconstructing the full process would require dated source crosswalks, definitions for all 71 variables, processing scripts and records of coding decisions. Some source pages also change without a public revision history, which limits retrospective verification.

## 3. Description of the dataset

The COSMI Hub v1.0.0 snapshot is archived on Zenodo (Dorsch et al., 2026; https://doi.org/10.5281/zenodo.23233124). The UTF-8 file cosmihub_master_2026-10-08.csv, released on 8 October 2026, contains 1,186 records and 71 variables. Zenodo records the MD5 checksum ca754336562d01eb63e16814afd086fe. We release the dataset under Creative Commons Attribution 4.0 International (CC BY 4.0).

The 71 variables describe names and identifiers, geography, provenance, record level, operational status, thematic activities, governance, funding and data quality. Table 1 lists the ten thematic variables, while Table 2 explains selected fields useful for interpreting and grouping records. These tables summarise the dataset but are not a complete dictionary of every field and value.

**Table 2. Selected variables, their interpretation and recommended use in v1.0.0.**

| CSV variable(s) | Description / analytical use | Status / recommended use |
|---|---|---|
| id | Stable internal record identifier (INIT_XXXX); not an external registry ID. | Current internal key; preserve for joins. |
| original_category | Historical single-label categorisation, based on da Silveira et al. (2023); retained for traceability, not the preferred field for analysing multidimensional topical scope. | Historical field; prefer dim_* for topical analyses. |
| country | Country that established/leads the initiative. In the case of more than one country, the label international is assigned. | Geographic base, not necessarily reach. |
| iso3 / latitude / longitude | Geographic attribution and representative map coordinates; some project codes are not ISO countries (e.g. regional/international initiatives). | Representative points; check ISO/project codes. |
| governance_type_openit / community_governance | Organisational governance model and degree of community participation. | Organisational fields; avoid conflation. |
| object_type_openit | Source-informed functional type of the documented object (e.g. infrastructure/tool/platform, publishing/dissemination or policy/guideline), distinct | Functional type; not an algorithmic |
| | from record granularity and statistical clustering. | cluster. |
| lead_actor_type_openit | Type of institution or actor leading the initiative; a separate concept from its governance model. | Lead actor category; verify allowed values. |
| scope_openit | Recorded scope of activity; verify coded values before treating it as the geographic service area. | Scope descriptor; do not equate with country. |
| nonprofit / non_profit (outdated) | Two non-profit-status fields retained in v1.0.0; their differing definitions, value sets and provenance require a verified mapping before reuse. | Unreconciled fields; do not merge. |
| initiative_family | Groups related initiatives, services and deployments without implying a formal legal parent–child relationship; use alongside record_level for counting. | Family link; not a legal parent–child relation. |
| record_level | Granularity of the documented record (standalone initiative, service, deployment or umbrella organisation); distinct from object_type_openit, which captures functional type. | Key counting field; avoid double counting. |
| operational_status | Documented lifecycle status, including ongoing, pilot, planned, discontinued or unknown. | Many values unknown; analyse cautiously. |
| source / source_record_id | Indication of source and upstream record keys where available; *_LEGACY keys are generated by COSMI Hub. | Retain original and *_LEGACY references. |
| source_url / data_quality | Location of supporting sources and recorded assessment of data quality. | Supporting provenance; check availability. |
| funding_model / economic_confidence | Funding arrangement and confidence in the supporting evidence (high, medium, or low). | Interpret funding with confidence ratings. |
| ror_id / wikidata_id / tsosi_id / identifier_enriched_by | External identifiers and source of matching. | Optional IDs; verify the matched entity level. |

The archived CSV contains both nonprofit and non_profit; their different definitions, allowed values and provenance have not yet been established in a verified crosswalk. Neither should be treated as interchangeable with the other or with governance_type_openit. Codes such as 'NA', 'unknown', 'unclear', 'partial' and 'possible' also have field-specific meanings and should not automatically be recoded as 'no'. The geography fields combine ISO 3166-1 alpha-3 country codes with project-defined labels such as INT for international entries. Some map coordinates are representative points rather than headquarters locations. We recommend importing fields as text initially and documenting any recoding.

### 3.1. Field definitions and coding conventions

The Zenodo deposit consists of one CSV file cosmihub_master_2026-10-08.csv with a header row. The archived CSV combines historical fields with variables introduced during harmonisation. For topical categorisation, original_category retains the earlier single-label assignment for traceability, whereas the ten dim_* variables represent the subsequent multi-label approach. The v1.0.0 deposit does not include the source registry, detailed coding-decision log or a complete dictionary of the 71 variables. Table 1 lists the topical dimensions; Table 2 documents selected fields and their recommended interpretation. A versioned dictionary covering all variables, permitted values and provenance would improve reproducibility and support future schema cleanup. Before recoding, users should distinguish missing, uncertain and negative values, particularly in status, governance and software variables.

We maintain the working database in Google Sheets alongside controlled vocabularies, candidate lists and audit worksheets. The R Shiny dashboard uses views prepared from this evolving collection. Its counts, fields and operational statuses may therefore differ from those in the fixed Zenodo release. All figures reported in this article refer to the 8 October 2026 CSV.

## 4. Reuse potential and precautions

When analysing counts, we recommend checking record_level and initiative_family before treating rows as independent initiatives. A service and its umbrella organisation may be recorded separately. Geographic fields indicate a documented base or representative location, not necessarily the territory served; some coordinates mark a country or region rather than an office. Comparisons across countries also depend on source coverage, language and available documentation.

Before converting dim_* fields to binary indicators, users should state whether 'possible' entries are excluded, counted as positive or analysed separately, and check whether this choice affects the result. For example, within the 8 October 2026 CSV, let A be the set of records coded 'yes' for dim_open_access and B the set coded 'yes' for dim_open_data. Jaccard similarity, $J(A,B) = |A \cap B| / |A \cup B|$, measures overlap between these two subsets of record IDs; it does not compare two releases. The R Shiny dashboard presents maps, typologies and co-occurrence analyses based on the evolving working dataset, so its results may differ from those obtained using the archived CSV.

For reproducibility, we recommend importing the UTF-8 CSV with fields initially read as text, retaining id and source_record_id, documenting how missing and uncertain values are recoded, and recording any filters on record_level or operational_status. Analyses should cite the version-specific Zenodo DOI.

The dataset can also be used to examine overlaps between thematic dimensions, gaps in external identifier coverage and relationships between umbrella organisations and services. Comparing releases could show how the curated collection has changed, provided that newly identified records are distinguished from changes in the initiatives themselves. Neither counts within one release nor differences between versions directly measure national open science policy performance.

## 5. Data access, licensing and preservation

The archived dataset is available as cosmihub_master_2026-10-08.csv at https://doi.org/10.5281/zenodo.23233124 (Dorsch et al., 2026). It contains 1,186 records and 71 variables and is licensed under CC BY 4.0. The live dashboard is available at https://amcm.shinyapps.io/openit/. We link our methodological notes and original thematic codebook under Supplementary project documentation.

The version-specific DOI identifies the fixed 8 October 2026 dataset even if the dashboard changes. The MD5 checksum allows users to check that a downloaded file matches the archived copy. CC BY 4.0 permits redistribution and adaptation with attribution and an indication of changes. The source inventories remain separate resources, however, and may have different licences; anyone reusing them directly should check their terms and cite their creators. The dataset and this paper are separate research outputs and should be cited accordingly.

## 6. Software and documentation

We provide the R Shiny dashboard code at https://github.com/abdelghani-maddi/osinit (app.R). The repository covers the application and its analyses, not a complete record of manual data-curation decisions.

## 7. Ethics, personal data and transparency

### 7.1. Data protection and reuse rights

We compiled COSMI Hub exclusively from publicly accessible organisational directories, official websites, institutional documents and policy reports. The records describe initiatives, infrastructures, services and policy instruments; we did not collect data from human research participants. The dataset is released under CC BY 4.0. This licence does not override third-party rights or applicable data-protection obligations where publicly listed professional contacts are identifiable.

### 7.2. Declaration of generative-AI assistance

ChatGPT (OpenAI, GPT-6; October 2026) assisted with drafting, restructuring and editing selected passages in the abstracts, dataset metadata and Sections 1-8. The authors alone are responsible for all content and citations and for the manuscript's accuracy, scientific integrity.

## 8. Outlook

- Future releases could extend coverage and improve the documentation of variable definitions, provenance and changes between versions.
- A further step would be to invite initiative representatives, where feasible, to review their own records. Corrections should be documented and distinguished from genuinely new initiatives.
- Researchers, infrastructure providers and initiative representatives are invited to use the archived dataset, cite its version-specific DOI and suggest additions or corrections to the corresponding author. Proposed changes would be checked against the same documentary criteria as existing records.

## Acknowledgements

We thank colleagues in the OPENIT project, the OSMI community and the TSOSI team for assistance with identifier enrichment, and members of the open-science community who suggested additions and corrections.

## Competing Interests

The authors declare that they have no competing interests.

## Funding

This work was supported by the French National Research Agency (Agence Nationale de la Recherche, ANR) through the OPENIT project (ANR-24-RESO-0001).

## Author contributions (CRediT)

Author initials: ID = Isabelle Dorsch; MAS = Marc-André Simard; AM = Abdelghani Maddi. Contributions: Conceptualization (ID, MAS, AM); Data curation (ID, AM); Formal analysis (MAS, AM); Funding acquisition (MAS, AM); Investigation (ID, MAS, AM); Methodology (ID, MAS, AM); Project administration (ID, AM); Resources (AM); Software (AM); Supervision (ID, MAS, AM); Validation (ID, MAS, AM); Visualization (AM); Writing – original draft (ID, MAS, AM); Writing – review & editing (ID, MAS, AM).

## Supplementary project documentation

- About — project background
- Methodology — development and validation notes
- Classification codebook — thematic categories
- FAQ — dashboard guidance